\documentstyle[aps]{revtex}

\newcommand{\beq}{\begin{equation}}
\newcommand{\eeq}{\end{equation}}
\newcommand{\four} {  {}^{(4)}\kern-1pt  }
\newcommand{\ben}{\begin{eqnarray}}
\newcommand{\een}{\end{eqnarray}}

\catcode`@=11     
\def\eqalign#1{\null\,\vcenter{\openup\jot\m@th
  \ialign{\strut\hfil$\displaystyle{##}$&$\displaystyle{{}##}$\hfil
      \crcr#1\crcr}}\,}
\catcode`@=12   

\begin{document}

\title{
\begin{flushright}
astro-ph/0209232 \\
\end{flushright}
Kinematical solution of the UHE-cosmic-ray puzzle\\ without
a preferred class of inertial observers}
\author{{\bf Giovanni AMELINO-CAMELIA}}
\address{Dipart.~Fisica,
Univ.~Roma ``La Sapienza'', and INFN Sez.~Roma1\\
P.le Moro 2, 00185 Roma, Italy\\
E-mail: giovanni.amelino-camelia@roma1.infn.it}

\maketitle

\begin{abstract}
Among the possible explanations for the puzzling observations
of cosmic rays above the GZK cutoff there is growing interest
in the ones that represent kinematical solutions, based either
on general formulations of particle physics with small violations
of Lorentz symmetry or on a quantum-gravity-motivated
scheme for the breakup of Lorentz symmetry.
An unappealing aspect of these cosmic-ray-puzzle solutions
is that they
require the existence of a preferred class of inertial observers.
Here I propose a new kinematical solution of the cosmic-ray puzzle,
which does not require the existence of a preferred class of inertial
observers. My proposal is a new example of a type of relativistic theories,
the so-called ``doubly-special-relativity" theories,
which have already been studied extensively over the last two years.
The core ingredient of the proposal is a
deformation of Lorentz transformations in
which also the Planck scale $E_p$
(in addition to the speed-of-light scale $c$)
is described as an invariant.
Just like the introduction of the invariant $c$
requires a deformation of the Galileian transformations into the
Lorentz transformations, the introduction of the invariant $E_p$
requires a deformation of the Lorentz transformations,
but there is no special class of inertial observers.
The Pierre Auger Observatory and the GLAST space telescope should
play a key role in future developments of these investigations.
I also emphasize that the doubly-special-relativity
theory here proposed,
besides being the first one to provide a solution for the cosmic-ray
puzzle, is also the first one in which a natural description of
macroscopic bodies is achieved, and may find applications
in the context of a recently-proposed dark-energy scenario.
\end{abstract}

\bigskip
\medskip


\section{Introduction}
Over the last few years there has been a large number of studies
concerning the emerging ``cosmic-ray puzzle."
UHECRs (ultra-high-energy cosmic rays)
should interact with the Cosmic Microwave Background
Radiation (CMBR) and produce pions.
Based on the typical energies of CMBR photons,
and on a straightforward
special-relativistic analysis of the kinematics of the
process $p+\gamma \rightarrow p+\pi$ (assuming the UHECR is
a proton),
one is led to the conclusion that photopion production
should render observations of UHECRs
with $E > 5 {\cdot} 10^{19}$eV (the GZK cutoff)~\cite{GZK}
extremely unlikely.
Above $10^{20}$eV the cosmic-ray mean free path should be only of
a few Mpc, and there are no astrophysical
sources capable of accelerating particles to such energies within
a few tens of Mpc from us.
Still, more than a dozen UHECRs have been reported by AGASA
with nominal energies at or above $10^{20}$ eV\cite{AgaWat}.
It is plausible that these UHECRs, the highest-energy particles
we have access to, may be providing us a window on a new
realm of fundamental physics. Proposed solutions extend from the idea
that previously unknown particles play one or another role in
evading the GZK cutoff to the idea, which is the main focus
of the study here reported, that the description of these UHECRs
as protons coming from cosmological distances is correct but the
GZK analysis requires modifications due to
new (``quantum") properties of spacetime (and the associated
laws of kinematics).
Of course, the simplest (and perhaps most likely) explanation
is that these preliminary experimental results may be misleading:
that for example AGASA might have overestimated the energies of
the observed UHECRs.

For what concerns the experimental side
the situation will become much clearer in
the not-so-distant future: especially
the Pierre Auguer Observatory~\cite{auger}
should definitely clarify whether violations of the GZK limit
are a reality.
In the meantime it is legitimate to take as working assumption that
we are (or may be) confronted with a ``cosmic-ray puzzle".
In this paper I propose a new solution for the puzzle.
My solution can be seen as a third example of ``kinematical solution"
of the cosmic-ray puzzle. I describe as ``kinematical solutions"
the solutions that are based on the assumption
that the description of these UHECRs
as protons coming from cosmological distances is correct and that
the GZK analysis requires modifications due to
a failure of Lorentz symmetry in the realm of
the relevant photopion-production processes.
At first this possibility was examined at a purely phenomenological level:
in the studies \cite{gonza,colgla}
a general parametrization of possible deviations from Lorentz
symmetry was introduced, and choices of parameters that would
shift upward the GZK limit were found.
Then, motivated by the proposal put forward in Ref.~\cite{grbgac},
there has also been work~\cite{kifu,gactp} on
a specific idea and scheme\footnote{There appears to be a certain tendency in
the literature to confuse the proposals based on particle-physics
models with a general parametrization of possible deviations from Lorentz
symmetry with the proposals based on quantum-spacetime models
with deviations from Lorentz symmetry. For example, in Ref.~\cite{dirbe00}
the Coleman-Glashow model of Ref.~\cite{colgla}
is described as a model of Planck-scale deviations from Lorentz
symmetry, whereas in Ref.~\cite{colgla}
there is no mention of the Planck scale,
quantum gravity, and quantum spacetime.
A phenomenology of Planck-scale deviations from Lorentz symmetry
was only proposed later in Ref.~\cite{grbgac},
and found its way in the cosmic-ray-puzzle literature only
starting with the studies reported in Refs.~\cite{kifu,gactp}.}
for the breakup of Lorentz symmetry,
in which the GZK limit is shifted upward by Planck-scale effects,
reflecting some new quantum properties of spacetime.

While very different on many levels, the two previous kinematical solutions
of the cosmic-ray puzzle, the general particle-physics phenomenology
of Refs.~\cite{gonza,colgla} and the class of quantum-spacetime
phenomenological models
of Refs.~\cite{grbgac,kifu,gactp}, have a common feature: the assumed
deviations from Lorentz symmetry require the emergence of
a preferred class of inertial observers (often, without any conceivable
justification, identified with the natural frame for the description
of the CMBR).
The primary objective of the study I am here reporting is to show
that this feature, which may be perceived as unactractive by some
colleagues, is not a necessary ingredient of kinematical solutions
of the cosmic-ray puzzle: I will discuss a Planck-scale deformation
of Lorentz symmetry which evades the GZK limit kinematically, but
does not require a preferred class of inertial observers.

My scheme is a new example of a class of relativistic theories,
the so-called ``DSR" or ``doubly-special-relativity" theories,
which I proposed in Ref.~\cite{dsr1dsr2}, motivated by a certain
perspective on the quantum-gravity problem,
and have already been studied extensively
over the last two years~[11-20]
DSR theories are relativistic theories that describe the laws
of transformation between inertial observers in such a way that the
Planck scale, $E_p$, acquires a role completely analogous to the one of
the speed-of-light constant $c$: in DSR both $c$ and $E_p$ are
observer-independent scales of the transformation laws.
It is indeed plausible~\cite{dsr1dsr2,leedsr}
that we might discover the need for this type of relativistic theories
in ultra-high-energy physics just in the same sense
that we discovered the need for special relativity in the study of
processes involving high velocities.

In the specific type of DSR theory that I propose as a solution
of the cosmic-ray puzzle
there are no deviations from special relativity for massless particles,
while for particles with finite mass, $m$,
the deviations from special relativity
are negligble at low energies but become significant when the particle's
energy $E$ is large enough that $m/E < E/E_p$.
For a proton the condition $m/E < E/E_p$ requires $E > 3 {\cdot} 10^{18}eV$,
and therefore the GZK cutoff happens to occur when the deviations
from special relativity are predicted to be rather large.
In general I observe that the ratio $E/m$ sets the magnitude of the boost,
while $E/E_p$ will set the magnitude of Planck-scale effects, and that
in a large class of DSR theories one should expect that the
deviations from ordinary Lorentz symmetry become significant when $E/m$
is large enough to compensate for the smallness of $E/E_p$.

In the next section I review the previously-proposed mechanism,
based on Lorentz-symmetry breaking,
for a quantum-gravity solution of the cosmic-ray puzzle.
In Section~3 I give a brief review of results in DSR theories,
in which Lorentz symmetry is deformed rather than broken.
In Section~4 I propose my DSR-based solution of the cosmic-ray puzzle.
Section~5 offers some closing remarks.

\section{Previous Planck-scale solutions
of the cosmic-ray puzzle}
It is plausible that the unification of
general relativity and quantum mechanics will involve some sort
of spacetime quantization, {\it e.g.} noncommutative geometry or
spacetime discreteness.
If spacetime is fundamentally discrete or noncommutative then
this should in particular apply to
the spacetimes that low-energy probes perceive as Minkowski (flat
classical and continuous).
From the point of view of experimental tests a key point is that
spacetime discreteness or noncommutativity are likely to require
departures from Lorentz symmetry (see, {\it e.g.},
Refs.~\cite{majrue,hooftlorentz}).

As a starting point for a phenomenology aimed at exploring
this type of possible quantum-gravity effects,
Refs.~\cite{grbgac,polonpap} proposed the investigation of
a scenario predicting violations of Lorentz symmetry
set by the Planck scale and characterized by the emergence of
a modified dispersion relation of the type
\begin{equation}
m^2 = E^2 - \vec{p}^2 + f(E,\vec{p}^2;E_p)
\simeq E^2 -  c^2 \vec{p}^2
+ \eta \left({E \over E_p}\right)^{n}  E^2
~,
\label{displead}
\end{equation}
where $\eta$ is a dimensionless coefficient of order 1.
The function $f(E,\vec{p}^2;E_p)$ is only specified in leading order
in $1/E_p$ since anyway experiments could at best find some
evidence of the leading contribution due to $f$.

This type of quantum-gravity scenarios is seen in analogy with the deformed
dispersion relations that apply to the description of collective modes
(such as phonons) in certain materials, whose propagation satisfies a
relativistic dispersion relation only up to corrections governed by the
scale of atomic structure of the material.
Similar considerations apply to the decription of the propagation
light in water. Intuitively one can indeed attempt to introduce in quantum
gravity the (however vague) concept of ``spacetime foam" and spacetime foam
may affect particle propagation in a way that to some extent can be viewed
in analogy with the way that the presence of other media affects particle
propagation.

One of the phenomenologically significant consequences
of deformed dispersion relations of the type (\ref{displead})
is the associated deformation of the thresholds for
certain particle-production processes.
Let me illustrate this effect in the specific example of photopion
production, $p + \gamma \rightarrow p +\pi$, which is
relevant for the cosmic-ray puzzle.
The kinematic condition for photopion production
is obtained by imposing conservation of energy and momentum
and by applying the dispersion relation (which in this case is deformed).
Denoting with $E$ the energy of the incoming proton
and denoting with $\epsilon$ the energy of the incoming photon,
one finds that photopion production is only possible if
\begin{equation}
E \epsilon - \eta {E^{2+n} \over 4 E_p^n}
\left({m_{prot}^{1+n} + m_\pi^{1+n} \over (m_{prot} + m_\pi)^{1+n}}
-1 \right) \ge {(m_{prot} + m_\pi)^2 - m_{prot}^2 \over 4}
~.
\label{gactpthresh}
\end{equation}
The $E_p \rightarrow \infty$ limit, in which the second term on the
left-hand side is neglected, of course reproduces the standard
special-relativistic result for the photopion-production
threshold condition.
In the processes we study in laboratory experiments
the $E_p$-dependent term is completely negligible, because of the
large value of the scale $E_p$.
But in the analysis of the cosmic-ray puzzle
one must consider an incoming
proton of very high energy, of order $10^{20}eV$,
while the photon is a CMBR photon with typical energy
roughly of order $10^{-3}eV$, and therefore one finds that
$E \epsilon \sim 10^{17}eV^2$,
$E^{3}/E_p \epsilon \sim 10^{32}eV^2$,
$E^{4}/E_p^2 \epsilon \sim 10^{24}eV^2$.
From these estimates one concludes~\cite{kifu,gactp}
that, for $n=1$ and $n=2$,
the formula (\ref{gactpthresh}) predicts a significant
shift upward of the GZK cutoff, and provides a solution of
the cosmic-ray puzzle.

This striking observation has motivated several follow-up
quantum-gravity studies based on the proposal (\ref{gactpthresh}).
More recently, other forms of Planck-scale modifications
of the dispersion are also being considered~\cite{mexicothresh}
in relation with the cosmic-ray puzzle.

These quantum-gravity-motivated solutions of the cosmic-ray puzzle,
just like the mentioned non-quantum-gravity solutions\cite{gonza,colgla}
that are also based on a breakup of Lorentz symmetry, necessarily require
the existence of a preferred class of inertial observers.
In fact, assuming that the Lorentz transformations maintain their
familiar form, the only invariant dispersion relation is $m^2 = E^2-p^2$.
All deviations from this special-relativistic form of the dispersion
relation will necessarily not be invariant: at least the values
of the coefficients that characterize the modified dispersion relation
(such as $\eta$ in (\ref{displead})) will be different for different
inertial observers and can be used to identify a preferred class of
inertial observers.
This is after all consistent with the mentioned
quantum-gravity intuition based on ``spacetime foam", which would indeed
be suitable for identifying a preferred class of inertial observers.

\section{Previous doubly-special-relativity theories}
The fact that the quantum-gravity scenarios mentioned in the preceding
section require a preferred class of inertial observers is not
necessarily a disappointment. As mentioned certain quantum-gravity ideas,
notably some perspectives on the spacetime-foam picture,
can provide motivation for exploring this possibility.
Moreover, in the study of particle physics the concept of spontaneous
breaking of a symmetry has proven very useful, and it is conceivable that
quantum gravity would host a similar mechanism for the spontaneous
breaking of Lorentz symmetry at the Planck scale.
For example, in string theory, the most popular
approach to the quantum-gravity problem,
mechanisms for the spontaneous braeking of Lorentz symmetry
have been extensively investigated
(see, {\it e.g.}, Ref.~\cite{stringSSBlorentz}).

However, the existence of a preferred class of inertial observers
is anyway not simple conceptually.
At present we can only make wild guesses about this
preferred class of observers.
Many studies attempt to
identify this preferred class
of inertial observers with a natural class of observers for
the CMBR, but this conjecture (although it cannot be excluded a priori)
appears to lack any justification, since there is no connection
between the classical physics responsible for CMBR physics
and the Planck-scale realm.
Moreover, in any case, even if one is not troubled by the possibility
of a preferred class of inertial observers,
it is natural to wonder~\cite{dsr1dsr2} whether
there are other alternatives in addition to the cases in which
Lorentz symmetry is exactly preserved at the Planck scale and the
case in which Lorentz symmetry is broken, with associated
emergence of a preferred class of inertial observers.
The so-called ``doubly-special relativity" (``DSR")
theories that I proposed
in Ref.~\cite{dsr1dsr2} are scenarios in which ordinary Lorentz
symmetry is not exactly preserved at the Planck scale, but
there is no special class of inertial observers.
Using a terminology which is popular in the mathematics community,
my proposal can be described as a ``deformation"
of Lorentz symmetry, without any true loss of symmetry.
This proposal can be motivated~\cite{dsr1dsr2} from the observation
that some quantum-gravity scenarios, notably a certain type
of scenarios based on noncommutative geometry, appear to require it.
It has also been conjectured~\cite{leedsr}
that DSR theories may be applicable to the popular ``loop quantum gravity"
theory.

Here, since I am focusing on the cosmic-ray puzzle, I can motivate
DSR theories following an analogy with the developments which
led to the replacement of Galileian relativity with special relativity.
In Galileian relativity there is no observer-independent scale,
and in fact the dispersion relation is written
as $E=p^2/(2m)$ (whose structure fulfills the requirements
of dimensional analysis without the need for dimensionful
coefficients). As experimental evidence in favour of Maxwell equations
started to grow, the fact that those equations involve a
special velocity scale appeared to require (assuming the Galilei
symmetry group should remain unaffected) the introduction
of a preferred class of inertial observers (the ``ether").
However, in the end we discovered that the Maxwell theory
does not require a preferred class
of inertial observers.
It was wrong to assume that Galileian relativity should apply
in all regimes. In the high-velocity regime it must be replaced
by special relativity.
Special relativity introduces the first observer-independent scale,
the velocity scale $c$, and its dispersion relation
takes the form $E^2 = c^2 p^2 + c^4 m^2$.
As interest in dispersion relations
of the type $c^4 m^2 =E^2 -  c^2 \vec{p}^2 + f(E,\vec{p}^2;E_p)$
is starting to grow within the quantum-gravity community,
also because of the analysis of cosmic-ray observations,
the fact that these dispersion relations involve a special energy
scale, $E_p$, is leading, as mentioned, to the assumption that
a preferred class of inertial observers might have to be introduced.
Could it be that this assumption is wrong?
Could it be that once again we do not need a preferred class
of inertial observers, but rather we need to deform once more the
laws of transformation between inertial observers?
DSR theories modify special relativity
in the exact same sense in which special
relativity modified Galileian relativity.

In Ref.~\cite{dsr1dsr2}, in proposing the idea of DSR theories,
I also derived the key features
of a first illustrative example (sometimes called ``DSR1")
of these theories.
This example,
whose key characteristic is the dispersion relation
\begin{equation}
2 E_p^2 \left[ \cosh ({E \over E_p}) - \cosh ({m \over E_p}) \right]
= \vec{p}^2 e^{E/E_p}
~,
\label{dispKpoin}
\end{equation}
has been analyzed in detail in a series of
papers~\cite{jurekdsr,jurekrossano,judesVisser,dsrdirac}
and is at this point reasonably well understood.

More recently, Maguejio and Smolin proposed~\cite{leedsr} a second
example of DSR theory, sometimes called ``DSR2",
whose key characteristic is the dispersion relation
\begin{equation}
m^2 = {E^2 - p^2 \over (1 - E_p)^2}
~,
\label{displee}
\end{equation}

For both of these theories we have now several results~[10-20]
The deformed Lorentz generators that implement
(\ref{dispKpoin}) (respectively (\ref{displee}))
as observer-independent laws have been explicitly contructed.
It has been established that the laws of composition of
energy-momentum (needed, for example, when we implement energy-momentum
conservation in a collision process) must be Planck-scale modified,
just in the same sense that the law of composition of velocity
(which is scale-independent in Galileian relativity) must be
speed-of-light-scale modified in special relativity.
A description of the deformed-boost action in terms of
a dependence of energy-momentum on
the rapidity parameter $\xi$ (the coefficient
of a boost generator $N$ in the exponentiation $e^{\xi N}$ that
implements a finite boost transformation)
has also been derived.

The relations involving rapidity are particularly useful
for a characterization of a DSR theory.
It is most convenient to focus on the amount
of rapidity needed to take a particle from its rest frame to a frame
in which its energy is $E$ (and its momentum is $p(E)$, which is
fixed, once $E$ is known, using the dispersion relation and the direction
of the boost). In the DSR1 theory one finds
\begin{equation}\label{xidsr1}
\cosh (\xi) = \frac{e^{E/E_p} - \cosh\left(m/E_p\right)}
  {\sinh\left( m/E_p \right)} ~,~~~
\sinh (\xi) = \frac{p e^{E/E_p}}
  {E_p \sinh\left(m/E_p\right)}\,\, ,
\end{equation}
while in DSR2 one finds
\begin{equation}\label{xidsr2}
\cosh (\xi) = \frac{E (1 - m/E_p)}{m(1 - E/E_p)} ~,~~~
\sinh (\xi) = \frac{p (1 - m/E_p)}{m(1 - E/E_p)} \,\, .
\end{equation}
Of course, in the $E_p^{-1} \rightarrow 0$ limit these relations
reproduce the corresponding special relativistic relations
\begin{equation}\label{xisre}
\cosh (\xi) = \frac{E}{m} ~,~~~
\end{equation}
\begin{equation}\label{xisrp}
\sinh (\xi) = \frac{p}{m} \,\, .
\end{equation}

These rapidity relations are extremely useful in the analysis
of DSR theories. On the basis of these relations one can introduce
some useful functions of the physical energy, momentum, mass of
a DSR theory. In DSR1 one introduces
\begin{equation}\label{auxdsr1}
\frac{{\cal E}(E,m)}{\mu(m)}
= \frac{e^{\lambda E} - \cosh\left(\lambda m\right)}
  {\sinh\left(\lambda m\right)} ~,~~~
\frac{{\cal P}(E,p,m)}{\mu(m)}
 = \frac{p e^{E/E_p}}
  {E_p \sinh\left(m/E_p\right)}\,\, ,
\end{equation}
while in DSR2 one introduces
\begin{equation}\label{auxdsr2}
\frac{{\cal E}(E,m)}{\mu(m)} = \frac{E (1 - m/E_p)}{m(1 - E/E_p)} ~,~~~
\frac{{\cal P}(E,p,m)}{\mu(m)} = \frac{p (1 - m/E_p)}{m(1 - E/E_p)} \,\, .
\end{equation}
The condition $\mu^2 = {\cal E}^2 - {\cal P}^2$ is understood
everywhere.
It is easy to show that in DSR1 and DSR2 the
functions ${\cal E}$ and ${\cal P}$ defined, as described, on the
basis of the rapidity relations, transform under the deformed
Lorentz boosts (which act on their arguments $E,p$)
just like special-relativistic energy and momentum would transform
under ordinary Lorentz boosts.
This allows to obtain most
results in DSR1 or DSR2 following a very simple
strategy: one does the analysis using
the auxiliary momenta ${\cal E}$ and ${\cal P}$,
which are easily handled because they transform under rotations
and boosts in the familiar way, and then the DSR result is obtained
by substituting in the final formula
${\cal E}$ and ${\cal P}$ with their expressions as
functions ${\cal P}(E,p,m)$, ${\cal E}(E,m)$ of the $E$ and $p$
which describe the physical energy and momentum in the DSR theory.

These useful properties of the functions ${\cal P}(E,p,m)$, ${\cal E}(E,m)$
are a consequence of the fact that the symmetries implemented in
the schemes DSR1 and DSR2 are actually a nonlinear realization
of the Lorentz symmetry group.
This simplicity is actually only present in the Lorentz sector:
studies of Poincar\'{e}-like symmetries that would extended the Lorentz
sectors of DSR1 and DSR2 are showing~\cite{jurekdsr,jurekDSRnew} that this
simplifications do not apply to the full Poincar\'{e} sector of the theory.
The simplifications only apply to the descriptions of Lorentz
transformations of energy-momentum space.
This is rather obvious from the perspective of applications of
the DSR framework in certain types of noncommutative spacetimes:
the fact that the spacetime coordinates do not commute introduces
(unless the commutators are themselves coordinate-independent)
complications for what concerns the description of the product
of plane waves (and the associated composition of momenta)
and the description of translations. Rotations and boosts are not
in {\it a priori} conflict with noncommutativity of the coordinates,
but they are affected by this noncommutativity through the fact that
their action of energy-momentum variables must reflect the
new properties of energy-momentum.

Looking further ahead in the research programme on DSR theories
one can conceive ways to introduce the second observer-indepedent
scale $E_p$ such that even the Lorentz sector is affected
by severe complexity. However, this possibility goes beyond the scope
of the present study. Here I intend to show that exactly a DSR
theory of the type described in this section, which in particular
relies on a nonlinear realization of the Lorentz group,
can be used to avoid the GZK cutoff.

\section{A new DSR theory that solves the cosmic-ray puzzle}
In Section~II I have discussed how quantum-gravity scenarios with
Lorentz-symmetry breaking can solve the cosmic-ray puzzle,
at the ``price" of introducing a preferred class of inertial observers.
In the previous section, Section~III, I have discussed the first
two examples, DSR1 and DSR2, of DSR theories.
Since the DSR framework does not introduce a preferred class of
inertial observers, if DSR1 and/or DSR2
were to solve the cosmic-ray puzzle,
the solution would indeed not require
a preferred class of inertial observers.
However, the analysis of
the threshold conditions for photopion production
in DSR1 and DSR2 does {\underline{not}}
lead to a solution of the cosmic-ray puzzle.
This analysis is actually very simple, because of the properties
described above of the functions ${\cal P}(E,p,m)$, ${\cal E}(E,m)$.
In DSR1 and DSR2 the special relativistic condition for photopion production
\begin{equation}
E \epsilon \ge {(m_{prot} + m_\pi)^2 - m_{prot}^2 \over 4}
~,
\label{secD1}
\end{equation}
is simply replaced by the condition
\begin{equation}
{\cal E}(E,m_{prot}) \epsilon \ge {[\mu(m_{prot}) + \mu(m_\pi)]^2
- [\mu(m_{prot})]^2 \over 4}
~.
\label{secD2}
\end{equation}
This result is valid to sufficient accuracy for the analysis of the
cosmic-ray puzzle; in particular,
for the energy of the CMBR photon I have not
even introduced the deformation function because CMBR photons
have such low energies that (as one can verify explicitly)
their description in DSR1 and DSR2 is basically identical to
the familiar special relativistic description.

As shown by Eqs.~(\ref{auxdsr1}),(\ref{auxdsr2}),
DSR1 and DSR2 do not give the same prediction for the form
of the function ${\cal E}(E,m_{prot})$, but in both cases
one can easily verify that the difference between (\ref{secD2})
and (\ref{secD1}) leads to a negligibly small modification of the GZK cutoff.

This observation about DSR1 and DSR2 has already been known
for some time~\cite{dsr1dsr2,frandar,leedsrnew}
and it has led to the expectation\footnote{Some authors~\cite{leedsrnew}
have given up so completely on the hope of finding a DSR solution for
the cosmic-ray puzzle that they arrived at proposing that solutions
of the cosmic-ray puzzle would necessarily require {\underline{two}}
length/energy scales, rather than the single one available in DSR theories.
The results I report in this paper provide counterexamples
to this conjecture.}
that DSR theories would in
general not provide solutions for the cosmic-ray puzzle, and therefore
a kinematical solution of the cosmic-ray puzzle would require
the emergence of a preferred class of inertial observers.

I will show here that instead there is a rather rich class of DSR theories
that does provide a solution to the cosmic-ray puzzle.
My analysis takes off from two related observations:

\noindent
$\bullet$ The data that presently are at the basis
of the cosmic-ray puzzle concern protons with energies in the range
from $\sim 10^{19}eV$ to $\sim 10^{20}eV$. Taking into account
that the proton has mass of $m_{prot}\sim 10^{9}eV$ (and the Planck scale
is of about $E_p \sim 10^{28}eV$) it is noteworthy that the paradox
emerges just above the scale $\sqrt{m_{prot} E_p} \sim 3 {\cdot} 10^{18}eV$.

\noindent
$\bullet$ The relativistic theories DSR1 and DSR2 are deformations
of ordinary special relativity which are governed by a single
dimensionless ratio, $E/E_p$, but this cannot be assumed as a generic
feature of DSR theories. Already in ordinary special relativity
the ratio $E/m$ plays a central role (see, {\it e.g.}, Eq.~(\ref{xisre})),
and it is therefore plausible (even natural) that in DSR theories
the scales $E/m$ and $E/E_p$ would combine in such a way that the
deformation would be also strongly characterized by the dimensionless
quantity $E^2/(m E_p)$.

From these two observations the careful reader will be already guessing
which type of DSR theories should be able to provide a solution
of the cosmic-ray puzzle: theories in which
the dimensionless quantity $E^2/(m E_p)$ is a strong characteristic of
the deformation and that are constructed in such a way that
for $E \ll \sqrt{m E_p}$ the effects induced by the deformation
are negligibly small, while for $E \ge \sqrt{m E_p}$
the effects are such that the GZK cutoff
is shifted significantly upward with respect to the prediction
of ordinary special relativity.

Even just within the simple type of DSR theory reviewed
in the previous section
(DSR theories in which the rotation/boost generators act on the
energy-momentum sector in a way that can be codified through
functions of the type ${\cal P}(E,p,m)$, ${\cal E}(E,m)$),
one can easily find deformations of the action of
rotation/boost generators on the energy-momentum space
that satisfy these conditions.
It is useful to focus on a couple of explicit examples.
Let me denote by DSR3a a DSR theory in which
the special-relativistic relation (\ref{xisre}) is replaced by
\begin{equation}\label{dispone}
\cosh (\xi) = {E \over m} - {E^3 \over m^2 E_p} \,\, .
\end{equation}
and denote by DSR3b a DSR theory in which (\ref{xisre}) is replaced by
\begin{equation}\label{dispgood}
\cosh (\xi) = {E \over m} (2 \pi)^{-E^2 \tanh[m^2 E_p^4/E^6]/(m E_p+E^2)} \,\, .
\end{equation}
These conditions define implicitly the functions  ${\cal E}(E,m)$
for DSR3a and DSR3b.
I don't even specify here the functions ${\cal P}(E,p,m)$,
since their implications for the GZK threshold are negligible.
The functions ${\cal P}(E,p,m)$ can be chosen, for example, in such a way
as to impose the condition that $c$ remains the maximum attainable
velocity.

Both in DSR3a, (\ref{dispone}), and in DSR3b, (\ref{dispgood}),
one finds an upward shift of the GZK cutoff
that is  large enough to provide a solution of the cosmic-ray puzzle.
In fact, in DSR3a the special-relativistic photopion-production threshold
condition (\ref{secD1}) is replaced by
\begin{equation}
\left( E - {E^3 \over m_{prot} E_p} \right) \epsilon \ge {(m_{prot}
+ m_\pi)^2 - m_{prot}^2 \over 4}
~,
\label{threA}
\end{equation}
in which I have neglected other correction terms which are subleading
with respect to the term $E^3/(m E_p)$,
and in DSR3b on finds that, for $10^{19}eV < E < 10^{21}eV$,
\begin{equation}
\left( E - {(2 \pi -1) E \over 2 \pi} \right) \epsilon \ge {(m_{prot}
+ m_\pi)^2 - m_{prot}^2 \over 4}
~,
\label{threB}
\end{equation}
in which I have omitted again some negligibly small $E_p$-dependent
corrections.
The data reported by AGASA provide support for an upward shift of
the GZK cutoff which is at least of a factor $6$. This shift is
clearly realized in (\ref{threB}), which predicts a shift by a factor $2 \pi$,
and in (\ref{threA}), which actually predicts that there cannot be any
photopion production involving CMBR photons (for $E > 10^{19}eV$
the quantity $E^2/(m_{prot} E_p)$ is greater than 1).

My two examples, DSR3a and DSR3b,
of DSR deformations with dependence on $E^2/(m E_p)$
only serve the purpose of illustrating some mechanisms by which
the dependence of the DSR deformation on $E^2/(m E_p)$
may lead to a relativistic theory that solves the cosmic-ray puzzle
(without a preferred class of inertial observers).
If this type of scenario is actually adopted by Nature it may
well make use of some other type of dependence on $E^2/(m E_p)$.

The example DSR3a was chosen in order to show that a deformation
with very simple structure can do the required task. A natural modification
of the DSR3a scheme could be obtained by adopting a deformation
function such that
\begin{equation}\label{disponegeneral}
\cosh (\xi) = f(E,m;E_p) \simeq {E \over m} - {E^3 \over m^2 E_p} \,\, ,
\end{equation}
where $f$ is a function such that the approximation described
in (\ref{disponegeneral}) holds at least
for $10^{19}eV < E < 10^{21}eV$.

The example DSR3b was chosen in order to show that a deformation
that leads to a solution of the cosmic-ray puzzle could also be
structured in such a way that the new effects are confined in a
relatively narrow range of scales.
The DSR3b scheme automatically leads to negligible effects
for massless particles (at all energies) while for particles with
finite mass $m$ the effects are nonnegligible only
for $\sqrt{m E_p} < E < (m E_p^2)^{1/3}$.

The (however {\it ad hoc}) deformation adopted in DSR3b
also has the property that, even if one assumed that these formulas
are applicable also to macroscopic bodies,
the structure of the deformation function is such
that the new effects are negligible for macroscopic bodies
(the effects vanish in the $m \rightarrow \infty$ limit).
In the general research programme on DSR theories a
key open problem~\cite{dsrnat}
concerns the DSR description of macroscopic bodies.
The available data on macroscopic bodies allow us to exclude
that deformations characterized by the dimensionless
quantity $E/E_p$ be applicable to these macroscopic bodies.
Since all DSR theories considered before the present paper were
based on $E/E_p$ deformations, a key assumption is that the deformation
only applies to microscopic particles. One can find~\cite{dsr1dsr2,dsrnat}
arguments to justify this assumption within DSR theories, but a
satisfactory description is still lacking.
When the deformation is governed, as here proposed,
by the dimensionless quantity $E^2/(m E_p)$ and if the deformation is
such that it can be neglected in the $E^2/(m E_p) \rightarrow \infty$
limit, as in the DSR3b case, then the deformation is automatically
confined to the realm of microscopic particles (for a macroscopic
body inevitably $E^2/(m E_p) \gg 1$), even before taking into account
the DSR arguments that may hint at some fundamental reasons for
confining the deformation to microscopic physics.

\section{Closing remarks}
I have shown here that there exist relativistic theories in which
the GZK cutoff is shifted significantly upward (enough to
explain the cosmic-ray puzzle) but there is no preferred class
of inertial observers. My proposal is a new type of DSR theory,
and it is therefore based on the idea that, just like a
speed-of-light deformation
of Galileian transformations turned out to be necessary when
we acquired the capability of exploring contexts involving ultra-high
velocities, we might need a Planck-scale
deformation of Lorentz transformations in order to describe data
obtained in ultra-high-energy contexts.

Of course, the fact that it is possible to find
a solution of the cosmic-ray puzzle of
the type here considered (kinematical, and without a preferred
class of inertial observers) should not in any way reduce interest
in the other kinematical solutions of the cosmic-ray puzzle.
My analysis shows that a kinematical solution is also possible
through a {\underline{deformation}} of Lorentz symmetry,
and therefore
without a preferred class of inertial observers, but the ultimate
answer will come from observations, and it may well be that Nature
hosts a kinematical mechanism of violations of the GZK limit,
based on a {\underline{violation}} of Lorentz symmetry,
and therefore requiring a preferred class of inertial observers.
If indeed Planck-scale physics is responsible for the violations
of the GZK limit, it is difficult to favour one or the other
scenario: we still know very little about quantum gravity, and
at present one appears to find equally plausible arguments that favour
a  Planck-scale violation of Lorentz symmetry~\cite{grbgac,gampul,thiem}
and arguments that favour
a  Planck-scale deformation of Lorentz symmetry~\cite{dsr1dsr2,leedsr}.

In this paper, which is mainly intended for the astrophysics community,
I have given a brief overview of DSR concepts and techniques
and I have described my new DSR proposal only to the extent that
is of interest for the cosmic-ray puzzle.
In a paper now in preparation~\cite{gacinprep}, which will be primarily
intended for the quantum-gravity community,
I will describe the proposal in detail,
working out all aspects of a couple of DSR theories with
deformation governed by the quantity $E^2/(m E_p)$.

The element of analysis provided here should be sufficient for
DSR novices to get started in the task of attempting to figure out
if there is a particularly compelling example of the type
of DSR theories here proposed, with deformation governed by
the dimensionless quantity  $E^2/(m E_p)$.
The deformations DSR3a and DSR3b that I discussed here should be seen
only as the starting point for these studies, since, of course, I cannot
argue in any way in favour of a special compellingness of those proposals.
I needed to formulate some specific proposals in order to illustrate the
fact that $E^2/(m E_p)$ deformations of DSR type are possible and that
they can provide solutions for the cosmic-ray puzzle.

While the search of a conceptually compelling proposal for
a $E^2/(m E_p)$ deformation of DSR type may prove insightful,
of course the ultimate task is to test the idea experimentally.
As discussed in detail
elsewhere~\cite{grbgac,kifu,gactp,dsr1dsr2,polonpap,billetal},
some planned observations can provide important input for
this type of research endeavor.
Clearly the key input will come from
the Pierre Auger Observatory~\cite{auger}, which should establish whether
or not there are indeed violations of the GZK cutoff and should
provide important hints concerning the possibility that the violations
be due to deformed kinematics.
Concerning the specific structure of the DSR deformation,
studies such as the ones planned by the GLAST space telescope~\cite{glast}
should provide important hints,
since they are sensitive~\cite{grbgac,dsr1dsr2,polonpap,billetal} to deformations
of the $E(p)$ dispersion relation, which is a key feature
of a DSR theory~\cite{dsr1dsr2,dsrnat}.

Among the possible avenues for extending the number of contexts
in which the class of DSR theories here proposed may find applications
it is worth mentioning, in closing, the scenario for dark energy
considered in Refs.~\cite{dark}. That scenario
is based on a specific assumption concerning the structure of the
dispersion relation. On the basis of the assumed deformation
of the dispersion relation it is also assumed that Lorentz invariance
should be broken at the Planck scale, with the associated
emergence of a preferred class of inertial observers.
The type of structure of the
dispersion relation which is adopted in these dark-energy scenarios
appears to be well suited for description within the type of
DSR theories here proposed.
If such a description is achieved one would
then have a dark-energy scenario with phenomenological
motivation analogous to the one advocated
in the studies~\cite{dark}, but without the
assumption of the existence of a preferred class of inertial observers.


\end{document}